\documentclass[slac_one]{revtex4}
\usepackage{graphicx}
\usepackage{fancyhdr}
\begin{document}

\title{Hadronic Event-Shape Variables at CMS} 

%

\author{G.\ Dissertori, F.\ Moortgat, M.\ Weber}
\affiliation{Institute for Particle Physics, ETH Zurich, 8093 Zurich, Switzerland}

\begin{abstract}
In this note a study of hadronic event shapes in
QCD events at the Large Hadron Collider (LHC) is presented.  
Calorimetric jet momenta, determined by various jet clustering algorithms, 
are used as input for calculating various event-shape variables which probe the structure of the hadronic final state. 
It is shown that the normalized event-shape distributions are robust
under variations of the jet energy scale and resolution effects, which makes them particularly suitable for early data analysis and tuning of Monte Carlo models. 
\end{abstract}

\maketitle

\thispagestyle{fancy}


\section{EVENT-SHAPE VARIABLES} 
\label{sec:theVariables}

Event shapes belong to the most widely used variables to study QCD dynamics, especially at $\mathrm{e^{+}e^{-}}$ and $\mathrm{ep}$ colliders. Event-shape observables are defined in terms of the four-momenta of jets in the final state.  Recently a large set of new hadronic event-shape variables 
has been proposed in Ref.\ \cite{Giuliaetal}. An important aspect of these variables
is their normalization to the total transverse momentum or energy in the event. Therefore it is anticipated that energy scale uncertainties should cancel out to a large extent. Thus
we believe that they represent an useful tool for very early measurements of 
the properties of QCD events at LHC and the tuning of Monte Carlo models. 


Analogously to the $\mathrm{e^{+}e^{-}}$ event shapes, one can define hadronic event shapes in the transverse plane, for example the central transverse thrust:
\begin{equation}\label{eq:centrthr}
T_{\perp,\mathcal{C}}\equiv \max_{\vec{n}_{T}}\frac{\sum_{i\in\mathcal{C}}\left|\vec{p}_{\perp,i}\cdot \vec{n}_{T} \right|}{\sum_{i\in\mathcal{C}}p_{\perp,i}}.
\end{equation}
where $p_{\perp,i}$ are the transverse momenta with respect to the beam axis $\vec{n}_{B}$. The transverse axis, for which the maximum is obtained, is the thrust axis $\vec{n}_{\text{T},\mathcal{C}}$. The variable which is typically used for perturbative calculations is $\tau_{\perp,\mathcal{C}}\equiv1-T_{\perp,\mathcal{C}}$. 
The central thrust minor is a measure for the out-of-plane momentum:
\begin{equation}\label{eq:centhrmin}
T_{m,\mathcal{C}}\equiv \frac{\sum_{i\in\mathcal{C}}\left|{p}_{x,i}\right|}{\sum_{i\in\mathcal{C}} p_{\perp,i}}= \frac{\sum_{i\in\mathcal{C}}\left|(\vec{p}\times\vec{n}_{B})\times\vec{n}_{\text{T},\mathcal{C}}\right|}{\sum_{i\in\mathcal{C}} p_{\perp,i}}.
\end{equation}

%


Below the results of a first simulation study \cite{CMS_PAS}  of these event-shapes variables at the Compact Muon Solenoid (CMS) are summarized.

\section{MONTE CARLO SAMPLES AND PRIMARY SELECTION}
$\text{PYTHIA\ 6.409}$ is used to simulate proton-proton collisions with a centre of mass energy $\sqrt{s}=14\,\text{TeV}$ \cite{Pythia}. 
The events have been passed through a full GEANT based simulation of the CMS detector. 
Events are preselected by requiring two or more calorimeter jets, corrected in their relative and absolute response, with a transverse energy $E_{\text{T}}\geq60\,\text{GeV}$ within a region of $|\eta|<\eta_{\mathrm{C}}=1.3$. 
If the two hardest jets of the event are not in this central region, the event is rejected. Only central corrected calorimeter jets with $E_{\text{T}}\geq60\,\text{GeV}$ are used for the event-shape calculation. The threshold on the transverse energy of the leading jet is set at $E_{\text{T},1}>80\,\text{GeV}$.  

\section{SENSITIVITY STUDIES AND SYSTEMATIC UNCERTAINTIES}
\label{sec:studies_sensi_syst}

\label{sec:corrections}

\begin{figure}[htbp!]
\begin{minipage}[t]{0.45\textwidth}
\centering 
\includegraphics[width=0.85\textwidth]{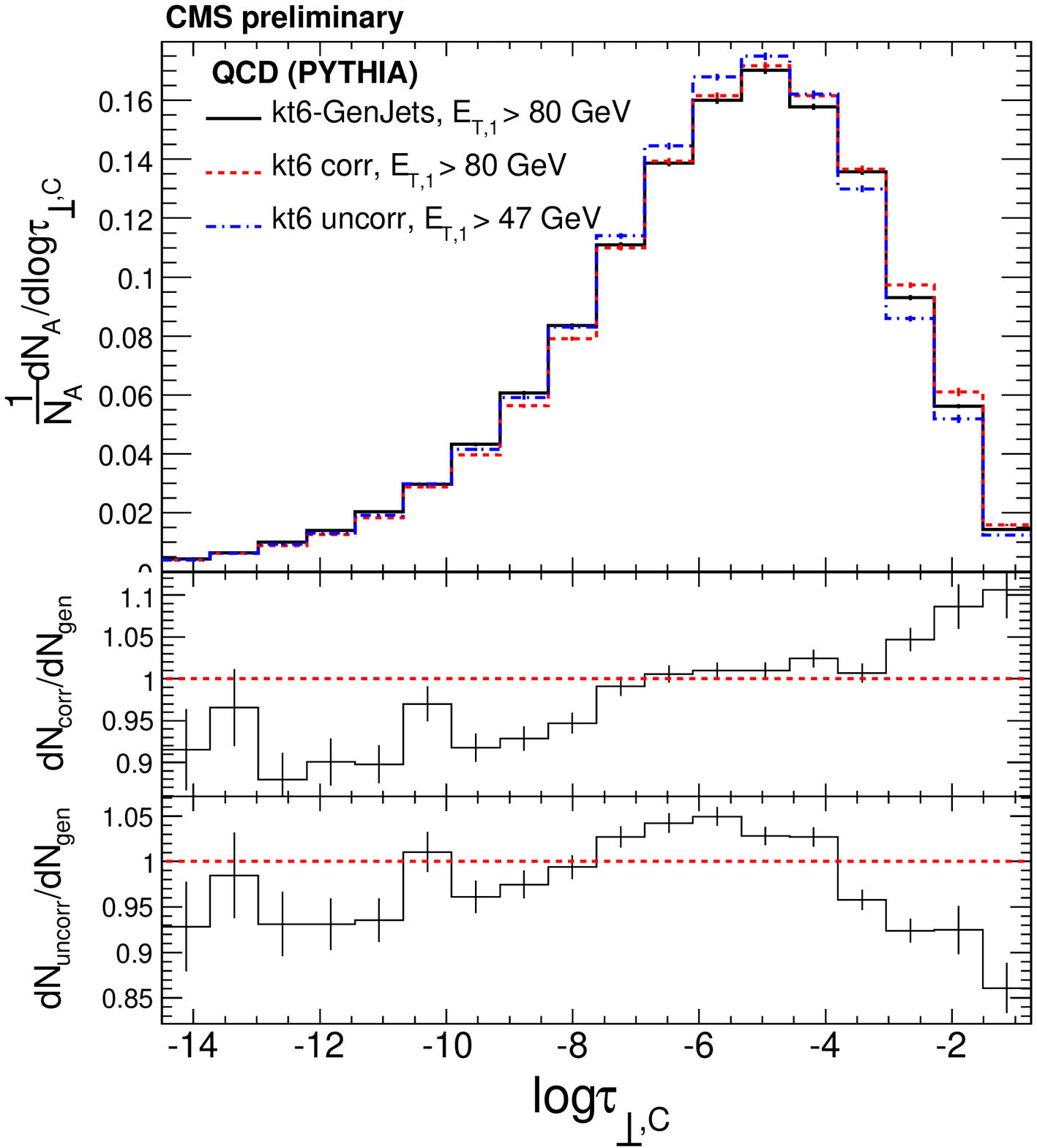}
\end{minipage}
\hspace{0.5 in}
\begin{minipage}[t]{0.45\textwidth}
\centering\includegraphics[width=0.85\textwidth]{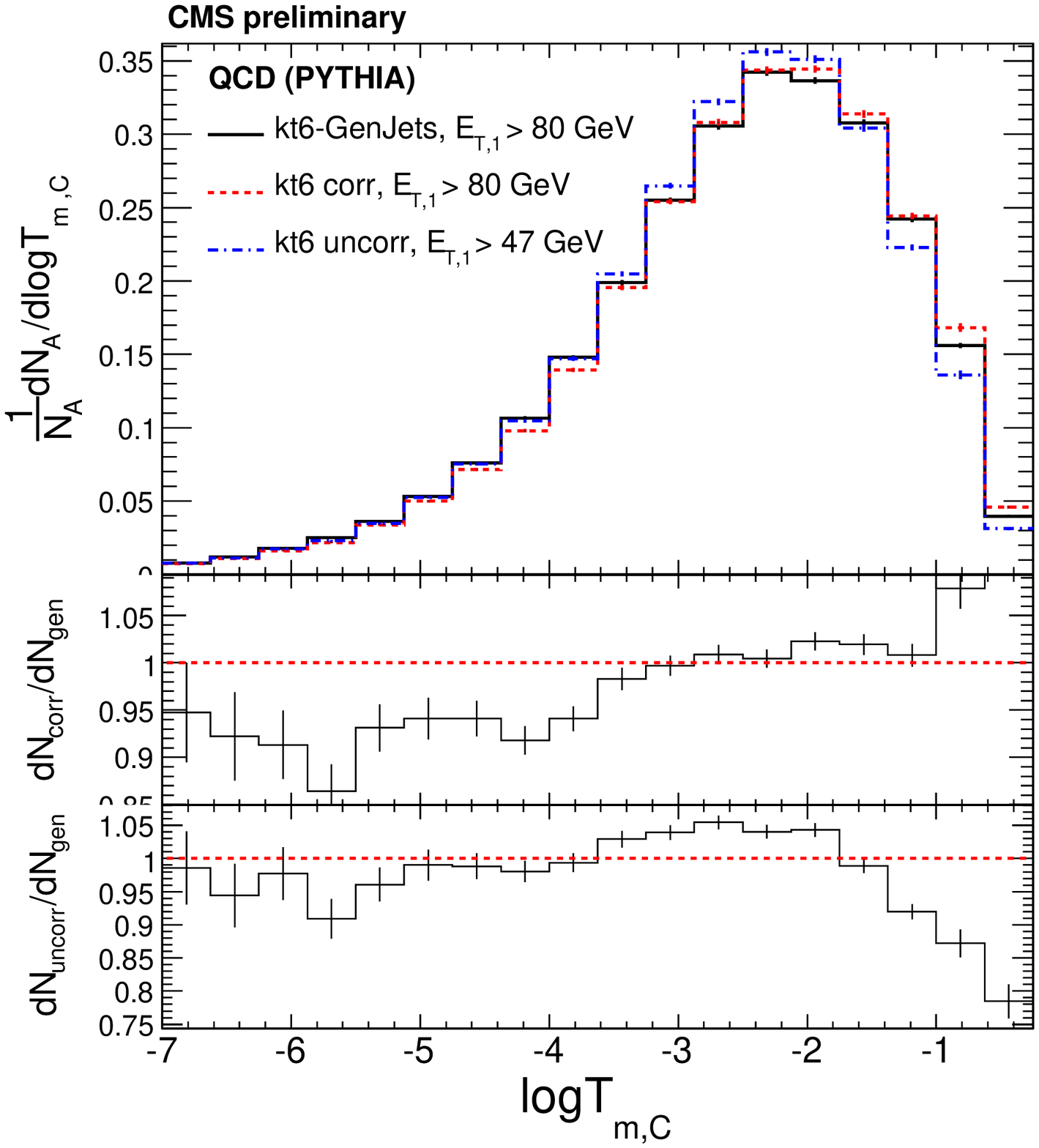}
\end{minipage}
  \label{fig:L2L3_thrust} 
   \caption{The effect of jet energy corrections on the central transverse thrust (left) and central thrust minor (right) distribution. The two lower histogram show the deviations to the generator level distribution.}
  \end{figure}

The effect of jet energy corrections on the normalized event-shape distributions can be judged by comparing the corrected and uncorrected distributions with the corresponding generator level distribution. A threshold of 30 $\text{GeV}$ on the transverse energy of uncorrected jets is applied, which corresponds to a generator level jet threshold of approximately 60 $\text{GeV}$. Similarly, the threshold on the uncorrected leading jets is $E_{\text{T},1}^{\text{uncorr}}>47\,\text{GeV}$. All three distributions agree well with deviations within 5-7 \% over most of the range as Fig~\ref{fig:L2L3_thrust} 
illustrates.

Often, the leading source of systematic errors in QCD data analysis is the limited knowledge of the jet energy scale (JES) and, to a lesser extent, the jet energy resolution. By definition, event-shape variables are expected to be rather robust against both sources of systematic errors. 
We assume a global uncertainty of 10\% on the knowledge of the jet energy scale. 
\begin{figure}[htbp!]
\begin{minipage}[c]{0.50\textwidth}
\centering 
\includegraphics[width=0.88\textwidth]{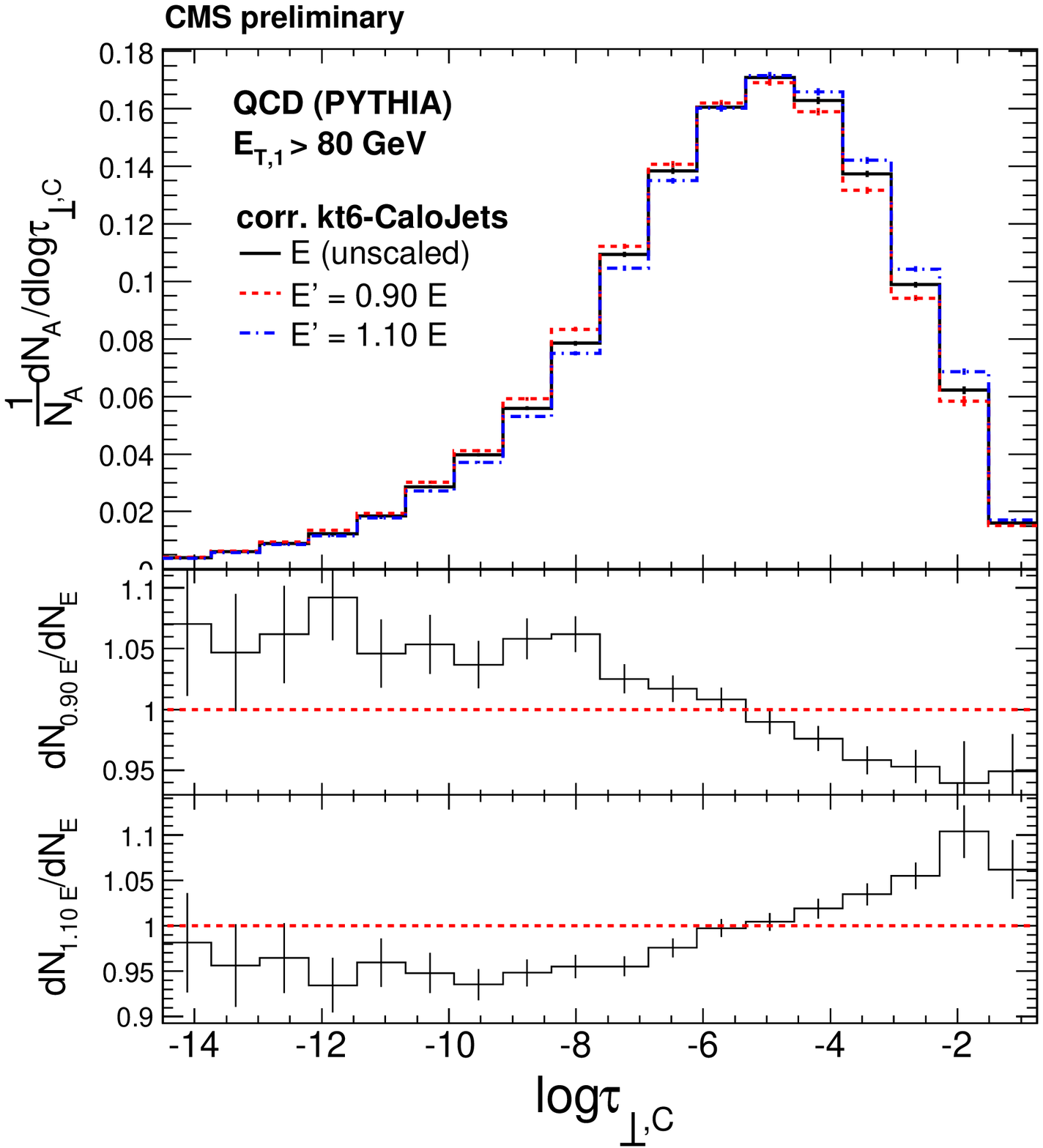}
\end{minipage}%
\begin{minipage}[c]{0.50\textwidth}
\centering \includegraphics[width=0.88\textwidth]{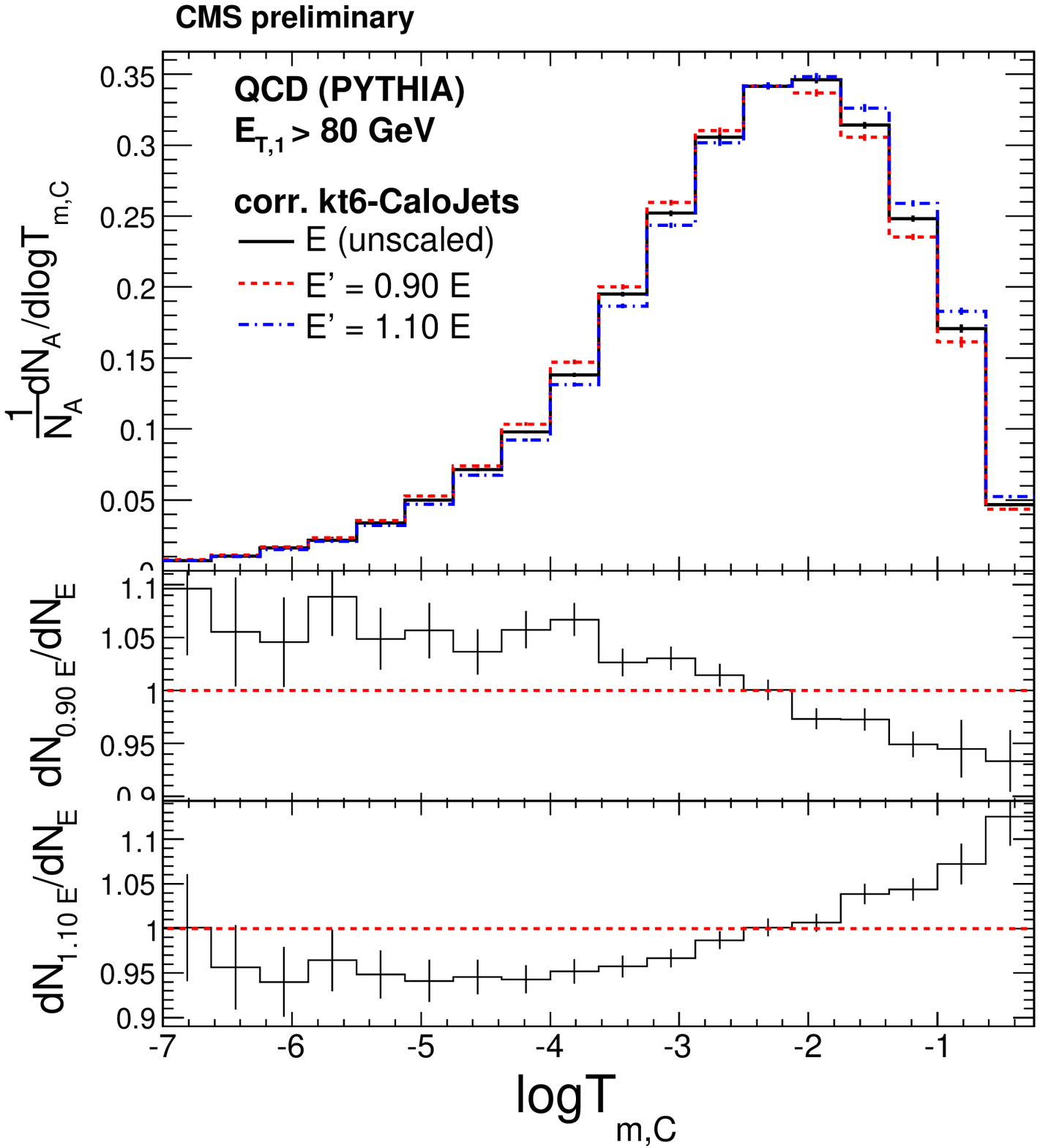}
\end{minipage}
\caption{The effect of a variation by 10\% in the jet energy scale on the transverse thrust (left) and central thrust minor (right) distribution using corrected kt6-Calojets. The two lower histograms show the deviations of the scaled distributions to the reference distribution.}
\label{fig:calo_thrust}
\end{figure}The resulting normalized event-shape distributions deviate by 5-10\% from the reference distribution over the whole energy range as can be seen in Fig.~\ref{fig:calo_thrust}.

The effect of the jet energy resolution is studied by applying the jet energy resolution smearing function \ref{eq:smear_ass} 
on generator level jets:
\begin{equation}
\frac{ \sigma( p_{\text{T}} ) }{ p_{\text{T}} } = \sqrt{ \left( \frac{5.2}{ p_{\text{T}} }\right )^2 + \left(\frac{1.2}{\sqrt{ p_{\text{T}} } }\right)^2 + (0.043)^2}
\label{eq:smear_ass}
\end{equation}

\begin{figure}[htbp!]
\begin{minipage}[t]{0.50\textwidth}
\centering 
\centering \includegraphics[width=0.88\textwidth]{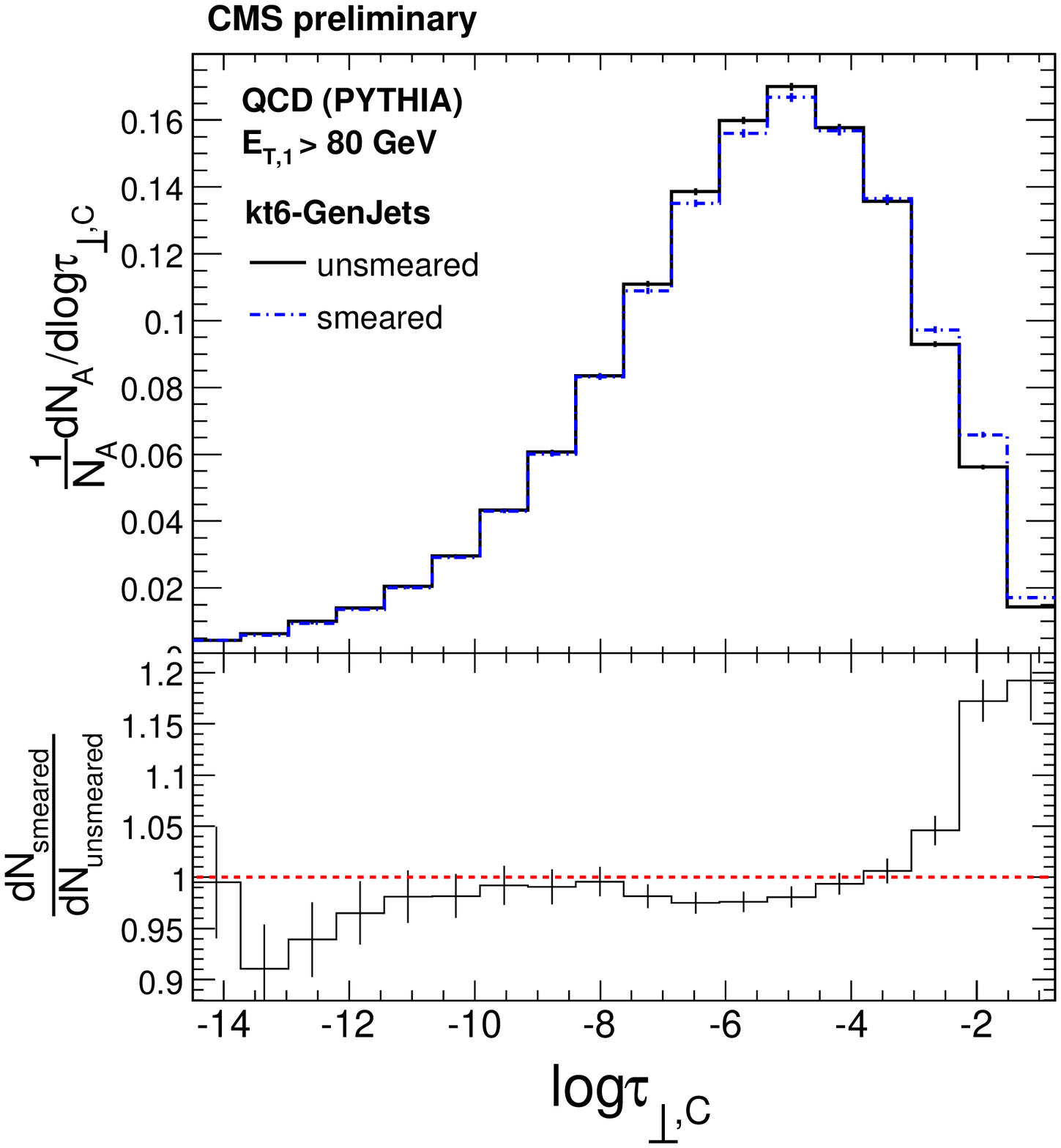}
\end{minipage}%
\begin{minipage}[t]{0.50\textwidth}
\centering \includegraphics[width=0.88\textwidth]{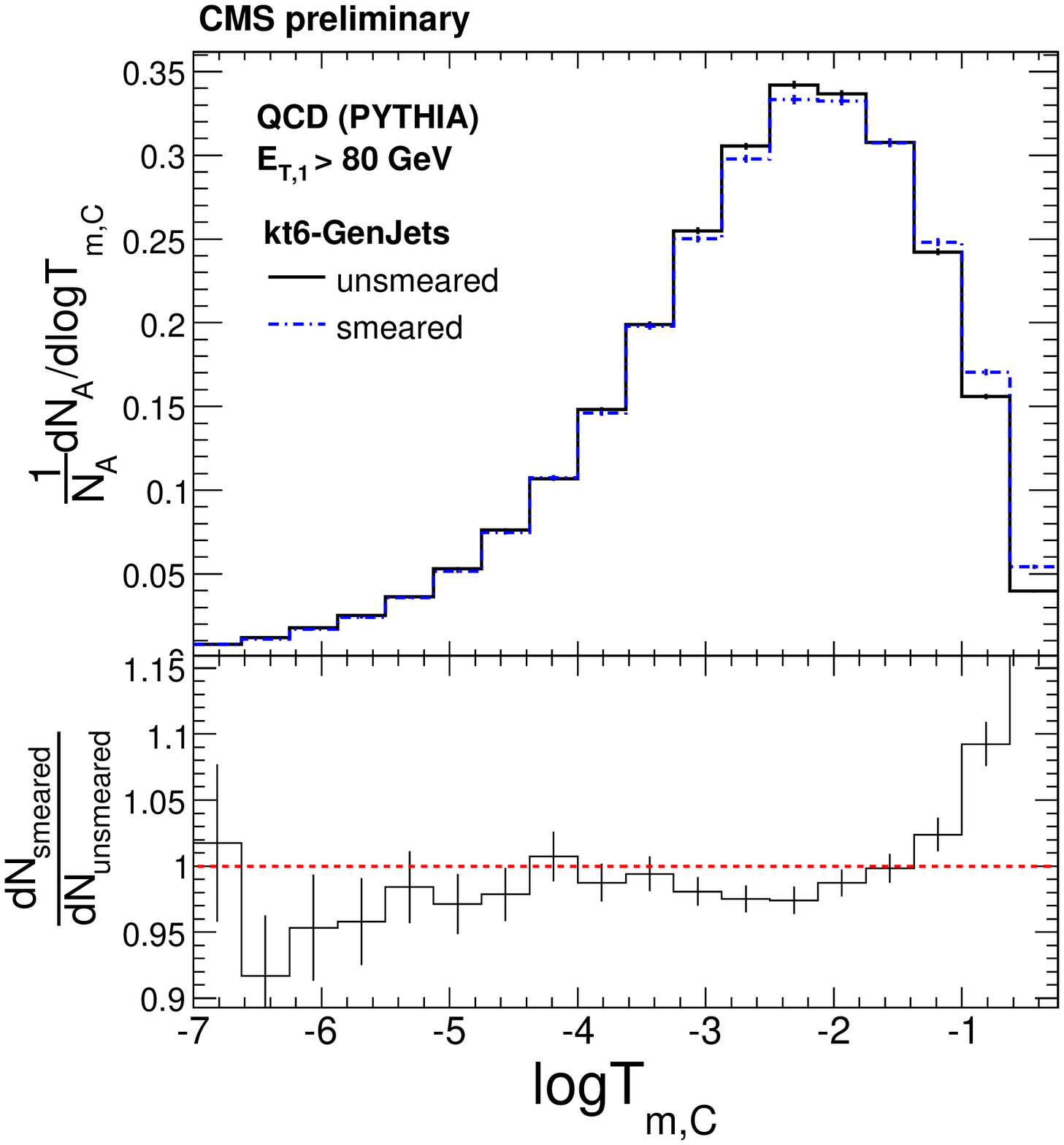}
\end{minipage}
  \caption{The effect of the jet energy resolution on the central transverse thrust (left) and central thrust minor (right). The lower histogram shows the deviation from the unsmeared distribution.}
\label{fig:smear} 
 \end{figure}
The smeared event-shape distributions deviate by less than $5\%$ from the unsmeared distribution over most of the energy range (Fig.~\ref{fig:smear}).



\section{RESULTS}
\label{sec:results}

In order to demonstrate the sensitivity of hadronic event-shape distributions to different models of multi-jet production, we compare the central transverse thrust and thrust minor distributions 
to the generator level predictions as obtained from two generators that contain different models of QCD multi-jet production, $\text{PYTHIA\ 6.409}$ and $\text{ALPGEN\ 2.12}$ \cite{Alpgen}. 
The $\text{ALPGEN}$ samples used in our study contains QCD processes from 2 up to 6 jets. 

\begin{figure}
\begin{minipage}[c]{0.50\textwidth}
\centering \includegraphics[width=0.9\textwidth]{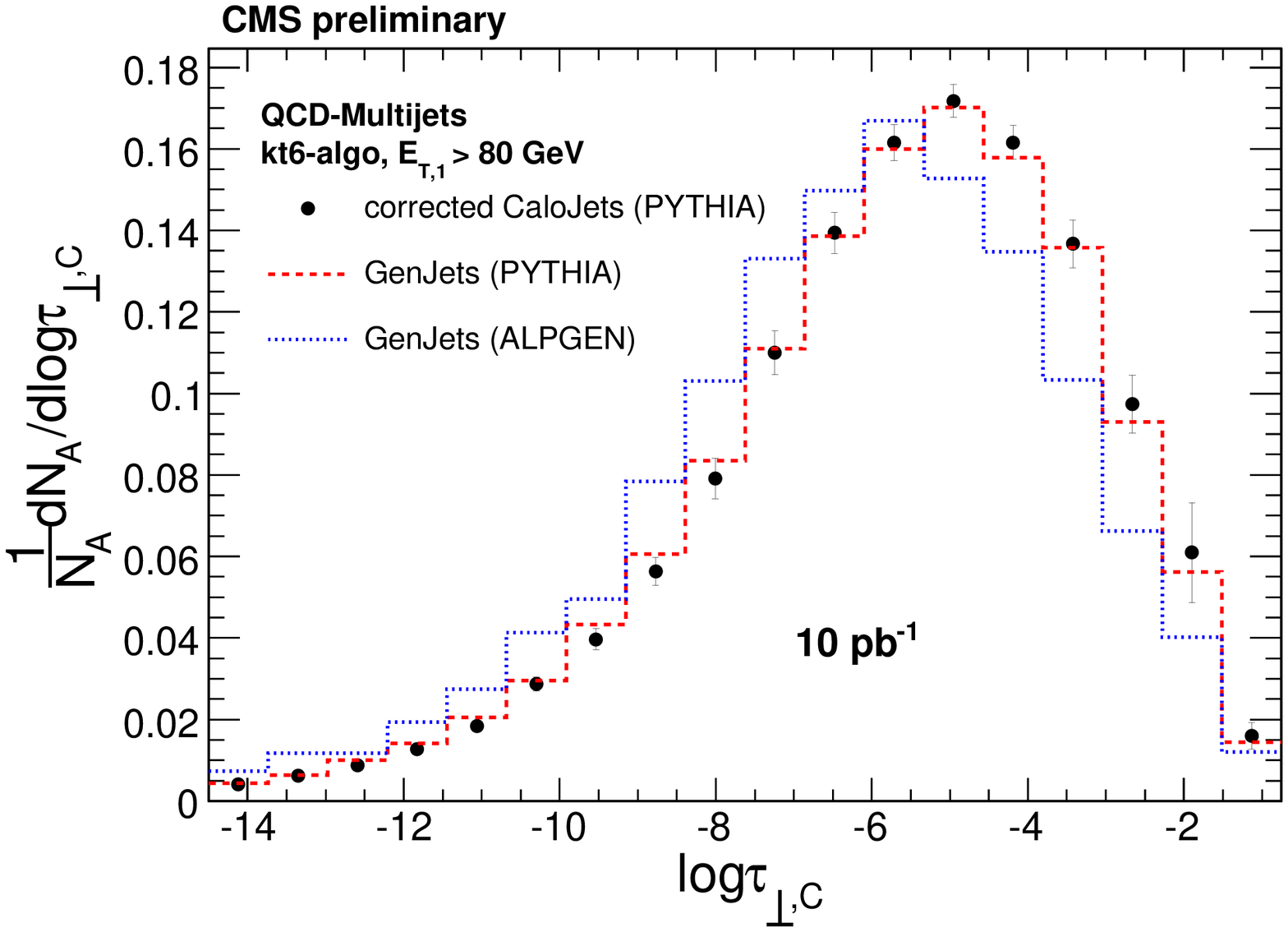}
\end{minipage}%
\begin{minipage}[c]{0.50\textwidth}
\centering \includegraphics[width=0.9\textwidth]{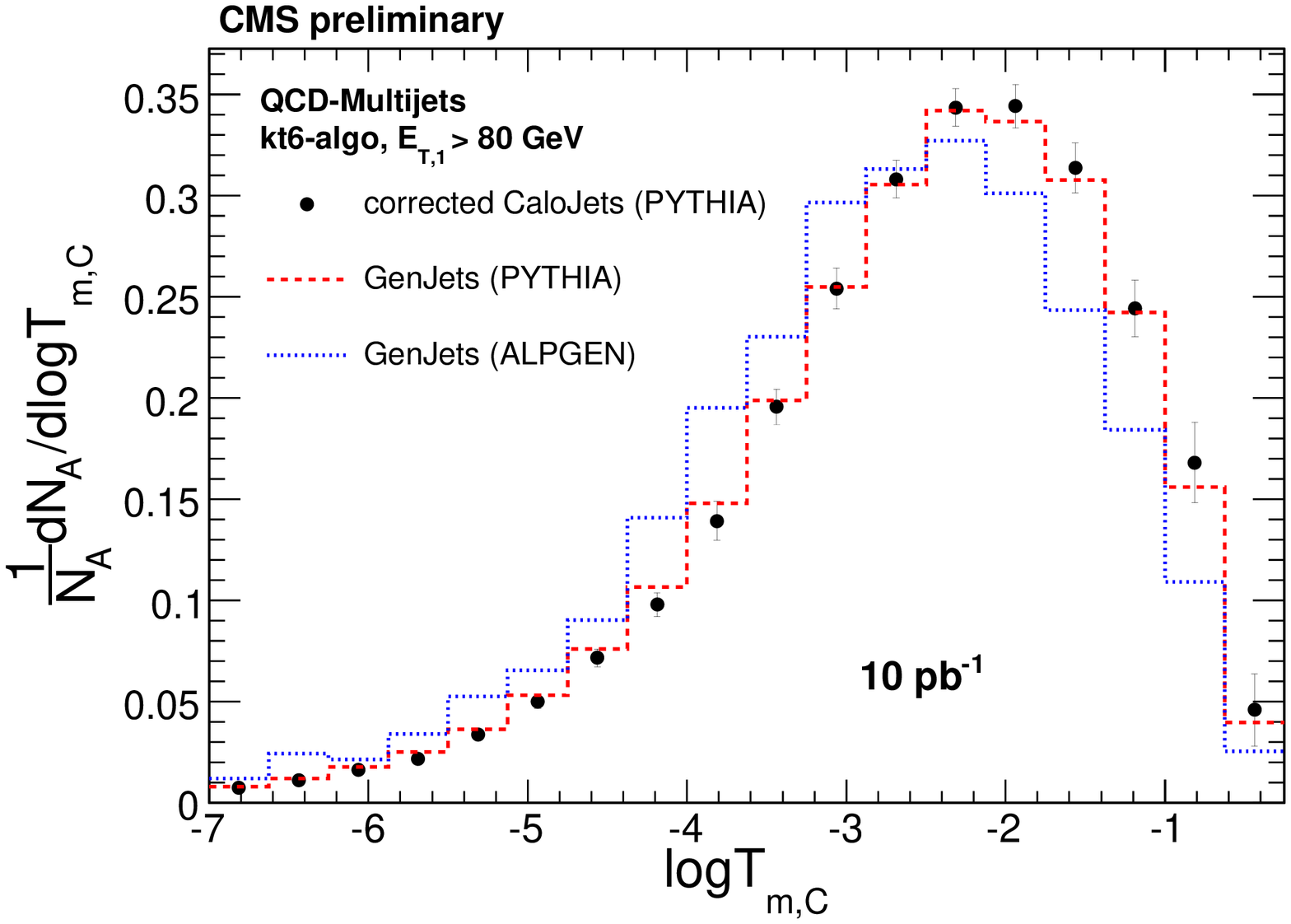}
\end{minipage}
\caption{The central transverse thrust (left) and the central thrust minor (right) distributions for $E_{\text{T},1} ^{\text{cor}} > 80\, \text{GeV}$ with the statistical and dominant systematic errors expected after $10\,\text{pb}^{-1}$ of integrated luminosity. The prescale of the trigger is assumed to be 100. The distributions are compared to the generator level distributions of $\text{PYTHIA}$ and $\text{ALPGEN}$. }
\label{fig:alp_pyt_trthr}
\end{figure}

In Fig.~\ref{fig:alp_pyt_trthr} the distributions of the central transverse thrust and central thrust minor can be seen. These events are selected from a jet trigger, based on the calibrated transverse energy of the hardest jet $E_{\text{T},1}> 60\,\text{GeV}$ with a prescale of 100. The error bars on the data points include the
statistical uncertainties corresponding to $10\,\text{pb}^{-1}$ of integrated luminosity and the systematic errors due to jet energy scale and jet energy resolution as discussed in the previous section. The corrected calorimeter jets correspond to the \text{PYTHIA} samples, and they are found to be compatible with the generator level jets from \text{PYTHIA}. It can be seen that there is a significant difference with respect to the \text{ALPGEN} distribution, reflecting the different underlying matrix element calculations in the generators and the different parameter choices.
The result shows that hadronic event shapes can be powerful handles in comparing and tuning different models of multi-jet production.

%




\section{CONCLUSIONS}
\label{sec:conclusions}

In this note we demonstrate the use of hadronic event shapes at the LHC. The event-shape variables are evaluated using calorimeter jet momenta as input. They are shown to be not very dependent on the effect of jet energy corrections. We present an estimate of the dominant systematic uncertainties at the startup, resulting from jet energy resolution effects and from the limited knowledge of the jet energy scale. Using the examples of the central transverse thrust and central thrust minor, we show that early measurements of event-shape variables allow to study differences in the modeling of QCD multi-jet production.

\begin{acknowledgments}
This research was supported in part by the Swiss National Science Foundation
(SNF).
\end{acknowledgments}

\end{document}